\documentclass[aps,prl,showpacs,twocolumn,reprint,superscriptaddress]{revtex4-1}
\usepackage{amsmath}
\usepackage{amssymb}
\usepackage{graphicx}
\usepackage{color}
\usepackage{multirow}
\usepackage{dcolumn}
\usepackage{textcomp}
\usepackage{longtable}
\usepackage{makecell}
\begin{document}

\title{Polarization multistates in antiferroelectric van der Waals materials}

\author {Guoliang Yu}
\affiliation{Key Laboratory for Matter Microstructure and Function of Hunan Province, Key Laboratory of Low-Dimensional Quantum Structures and Quantum Control of Ministry of Education, Synergetic Innovation Center for Quantum Effects and Applications (SICQEA), School of Physics and Electronics, Hunan Normal University, Changsha 410081, China}

\author {Shengxian Li}
\affiliation{Key Laboratory for Matter Microstructure and Function of Hunan Province, Key Laboratory of Low-Dimensional Quantum Structures and Quantum Control of Ministry of Education, Synergetic Innovation Center for Quantum Effects and Applications (SICQEA), School of Physics and Electronics, Hunan Normal University, Changsha 410081, China}

\author {Anlian Pan}
\affiliation{Key Laboratory for Micro-Nano Physics and Technology of Hunan Province, College of Materials Science and Engineering, Hunan University, Changsha 410082, China}

\author {Mingxing Chen}
\email{mxchen@hunnu.edu.cn}
\affiliation{Key Laboratory for Matter Microstructure and Function of Hunan Province, Key Laboratory of Low-Dimensional Quantum Structures and Quantum Control of Ministry of Education, Synergetic Innovation Center for Quantum Effects and Applications (SICQEA), School of Physics and Electronics, Hunan Normal University, Changsha 410081, China}
\date{\today}

\author {Zhenyu Zhang}
\affiliation{International Center for Quantum Design of Functional Materials (ICQD),
Hefei National Laboratory for Physical Sciences at Microscale, and
Synergetic Innovation Center of Quantum Information and Quantum Physics,
University of Science and Technology of China, Hefei, Anhui 230026, China}

\begin{abstract}
The bistability of charge polarization in ferroelectric materials has long been the basis of ferroelectric devices. However, the ferroelectricity
tends to be vanishing as the thickness of materials is reduced to a few nanometers or thinner due to the depolarization field. Instead, they show a
paraelectric or an antiferroelectric ordering in the ultra-thin limit, which is unfavorable for their applications in devices. Here we uncover
polarization multistates in thin films of van der Waals materials, in which the individual monolayers have an antiferroelectric ordering with
out-of-plane polarizations. This property results from a unique combination of the polarization and layer degrees of freedom. Using first-principles calculations, we demonstrate that bilayers and trilayers of the CuInP$_2$S$_6$ family possess quintuple and septuple polarization states., respectively. Our
climbing image nudged elastic band calculations for the bilayers and trilayers of CuInP$_2$S$_6$ and CuCrP$_2$S$_6$ further show that the states can be transformed into each other under appropriate
external electric fields, for which a unique layer-selective half-layer-by-half-layer flipping mechanism governs the transformings. Our study
opens up a door to design unusual polarization states using intrinsic degrees of freedom of layered antiferroelectrics for the next-generation
ferroelectric devices that go beyond the bistability paradigm.
\end{abstract}

\keywords{Ferroelectricity; two-dimensional materials; van der Waals materials; multistate polarization}

\maketitle
Ferroelectric (FE) materials have two polarization states that can be switched by applying electric fields. This feature enables their applications
in nonvolatile memory devices with high-density data storage and low power consumption~\cite{muralt2000ferroelectric,setter2006ferroelectric,catalan2009physics,martin2016thin,vaz2021epitaxial}.
To further increase the storage density and continue miniaturization of the FE devices, the thickness of the FE materials have to be reduced.
Unfortunately, the ferroelectricity is vanishing as the thickness of conventional FE perovskites reduced down to a few nanometers due to the
depolarization field~\cite{Zhong1994,junquera2003,Fong2004,dawber2005physics,Almahmoud2010}.

Layered van der Waals (vdW) FE materials are promising in addressing this scaling issue due to the fact that they have no dangling bonds at the surfaces
and weak interlayer interactions. For this purpose, a large number of two-dimensional (2D) FE materials have been experimentally discovered and theoretically
predicted~\cite{chang2016discovery,fei2016ferroelectricity,higashitarumizu2020purely,chang2020microscopic,huang2018prediction,Xu2017,Wan2017,Lin2019,Zhong2019,
XuChangsong2020,Xu2020,Ma2021,Ding2017,Zhou2017,Cui2018,Liu2016,Belianinov2015,Song2017,Qi2018,Lai2019,Hua2021,Huang2021,Sun2019,Feng2020,guan2022electric}.
In particular, the group-IV monochalcogenide monolayers with in-plane polarizations exhibit much higher Curie temperatures than their bulk phases~\cite{chang2016discovery}.
Moreover, a few groups of 2D FE materials, e.g., $\alpha$-In$_2$Se$_3$ monolayer~\cite{Ding2017,Zhou2017,Cui2018} and CuInP$_2$S$_6$~\cite{Liu2016,Belianinov2015}
thin films, possess out-of-plane polarizations, which are desired for FE devices. The bistability of the out-of-plane polarizations in these systems also forms
the basis of FE tuning of electronic properties of materials. In heterostructures of the 2D FE materials, switching the polarization of the FE substrates can
drive magnetic and electronic transitions in the overlayers~\cite{gong2019multiferroicity,zhang2020heterobilayer,bai2020nonvolatile,chen2022ferroelectric,cheng2021nonvolatile,zhu2019electric,sun2020controlling,hu2020electrical}.
However, a large number of the CuInP$_2$S$_6$ analogs (hereafter referred to as ABP$_2$X$_6$) instead show an intra-layer antiferroelectric (AFE) ordering down
to the monolayer limit~\cite{Song2017,Qi2018,Lai2019,Hua2021,Huang2021,Sun2019,Feng2020}, hindering their applications.

vdW stacking turns out to be an effective way of manipulating FE properties of 2D materials. It was shown that stacking nonpolar monolayers could drive
their bilayers into a FE ordering with an out-of-plane polarization~\cite{li2017binary,fei2018ferroelectric,liu2020magnetoelectric,yasuda2021stacking,vizner2021interfacial,wang2022interfacial,wan2022room}.
In these systems, the polarization is switched via a mechanism of interlayer sliding. For WTe$_2$ trilayers, the FE transition is accompanied by a sign reversal
of Berry curvature and its dipole, enabling electrical control of topological properties~\cite{Wang2019,Xiao2020}. Moreover, our recent study revealed that such
a vdw-type stacking could tune the AFE ordering of some ABP$_2$X$_6$ monolayers into the FE ordering~\cite{yu2021interface}.

In this Letter, we find that stacking the ABP$_2$X$_6$ monolayers with the undesired AFE ordering  gives rise to unexpected switchable polarization
multistates that goes beyond the concept of bistability.  This strategy makes use of the polarization and layer degrees
of freedom (DoF). Furthermore, the intralayer AFE ordering favors a mechanism of layer-selective half-layer-by-half layer flipping for transforming
of the states. We demonstrate the concept in thin films of CuInP$_2$S$_6$ and CuCrP$_2$S$_6$ by performing first-principles calculations, which show
that the bilayers and trilayers possess quintuple and septuple states, respectively. Moreover, our climbing image nudged elastic band calculations reveal
that the unique combination of the polarization and layer DoF yields an increasing barrier gradient as a state transforming to the one further. Thus,
the states can be transformed into each other under appropriate electric fields.

\begin{figure}[h]
  \includegraphics[width=0.9\linewidth]{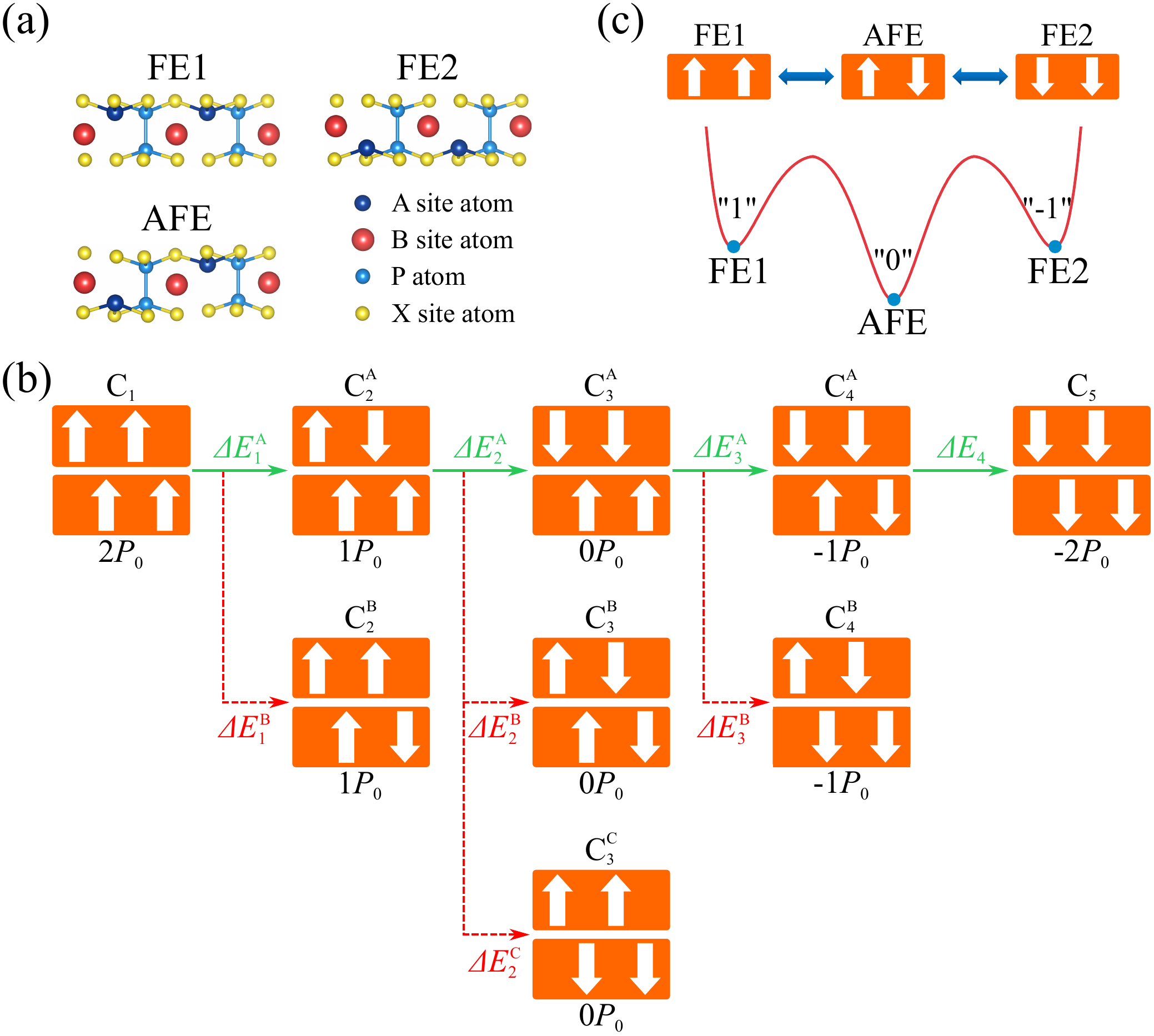}
\caption{Stacking-induced multiple polarization states in thin films made of AFE monolayers. (a) Polarization states in ABP$_2$X$_6$ monolayers. FE1 and FE2 represents
the two FE states with up and down polarizations, respectively. They are defined as the "1" and "-1" states, respectively. The AFE ordering is defined as the "0" state.
The paraelectric state is not shown here. (b) Quintuple polarization states in a ABP$_2$X$_6$ bilayer, which are classified by the value of $P_{tot}^{2L}$ in the unit
of $P_0$ (the polarization of the FE ABP$_2$X$_6$ monolayer). Configurations for each polarization state are shown in the same column. The arrows denote the polarizations.
The green lines denote the pathways that may have lower energy barriers than those denoted by the red lines as the system transforming from "2" to "-2". $\Delta E_i$ represent
the energy barriers for the transformings between neighboring states. (c) The kinetic pathway of transforming between the FE states in a ABP$_2$X$_6$ monolayer, for which
the AFE ordering is the ground state.}
 \label{fig1}
\end{figure}

We begin by presenting the concept of polarization multistates hidden in multilayers made of AFE monolayers. In an AFE monolayer like CuInP$_2$S$_6$ family (hereafter referred
to as ABP$_2$X$_6$), the out-of-plane displacements of the $A$ atoms are the main source of the intra-layer FE and AFE orderings (Fig.~\ref{fig1}a). For many of this family,
the AFE ordering are energetically lower than the FE ordering by about 10 - 80 meV\citep{yu2021interface}. For the AFE ordering, the unit cell is a 1 $\times$ 2 supercell of
the chemical primitive cell, in which half of the polarizations point along $z$ and the other half along -$z$ direction. In addition, we have also performed calculations for various stripe configurations  which have a higher energy than the AFE ordering. When two of them stack into a bilayer, the total charge
polarization, denoted as $P_{tot}^{2L}$ are expected to be approximately the superimpose of the monolayers because of the vdW-type interlayer interaction.  Indeed, our first-principles calculations confirm this expectation and show that the polarization saturates at thin films thicker than quintuple layers.  As a result,
$P_{tot}^{2L}$ can have quintuple values, i.e., $mP_0$ ($m$ = -2, -1, 0, 1, and 2), where $P_0$ is the polarization of the FE ordering of the monolayer. Fig.~\ref{fig1}b
schematically shows the polarization configurations for each type of $P_{tot}^{2L}$. For the "$\pm$ 2" states, there is only one configuration for each of them. For
the "$\pm$ 1" and "0" states, there are two and three inequivalent configurations, respectively. We denote the polarizations within each layer by arrows. A configuration
like $\frac{\uparrow \downarrow}{\downarrow \downarrow}$ refers to that the top layer is in the AFE ordering, whereas the bottom layer is in the FE ordering with polarizations
along -$z$ (see C$_4^B$ in Fig.~\ref{fig1}b). Likewise, there are septuple polarization states in trilayers (see the supporting information).

In the freestanding FE monolayers, the total energies of the FE states remain unchanged upon reversal of the polarization. However, for multilayers, flipping
the polarizations in different layers may give rise to different structures and symmetries. Thus, the energy of each configuration is strongly dependent on the
specific combination of the intra- and inter-layer polarizations. For instance, the energies of configurations $\frac{\downarrow \downarrow}{\uparrow \uparrow}$
and $\frac{\uparrow \uparrow}{\downarrow \downarrow}$ (C$_3^A$ and C$_3^C$, respectively, in Fig.~\ref{fig1}b) for bilayers should be different since their
structures are distinctly different in that in the former, the $A$ atoms are located close to the interface, whereas in the latter they are further away from it.
However, one can still expect that the lowest energy configuration is most likely among those with $P_{tot}^{2L}$ = 0 (the "0" state) due to the depolarization field.
There are three configurations for this state for the bilayers. The exact lowest-energy configuration and energy differences between the configurations should be
material-specific, in which the layer interaction plays an important role.

The diversity in configurations and energies of ABP$_2$X$_6$ multilayers results from the surface effect, which is vanishing as the thickness of the materials increases.
In the bulk phase of CuInP$_2$S$_6$ with the FE ordering being the ground state, there are two layers in the unit cell. The configuration with flipped polarizations in
the top layer ($\frac{\uparrow \downarrow}{\uparrow \uparrow}$) has the same energy with the one that has flipped polarizations in the bottom layer
($\frac{\uparrow \uparrow}{\uparrow \downarrow}$) due to the periodic boundary condition.

For freestanding ABP$_2$X$_6$ monolayers, the AFE ordering favors a half-layer-by-half-layer flipping of polarizations for the transforming of polarization states.
Despite the AFE state has a lower energy than the FE ones, it still can be transformed into the FE states under sufficiently large electrical fields. Assuming that
the monolayer is now in the FE1 ordering. One can expect two possible pathways for this state transformed into the other one, i.e., FE1-PE-FE2 and FE1-AFE-FE2. Due
to the fact of the AFE ordering being the ground state, the energy barrier for the pathway FE-AFE is lower than those for the pathways such as FE-PE and FE-stripe, which is confirmed by our
calculations for the AFE ABP$_2$X$_6$ monolayers. Therefore, for this kind of systems, the FE orderings will firstly transform into the AFE ordering (FE-AFE)
by an electric field that overcomes the barrier for the FE-AFE path. Then they will be driven from the AFE state into the other FE state (AFE-FE) by overcoming a larger
barrier than that for the FE-AFE transition (see Fig.~\ref{fig1}c).

For the bilayers, the top and bottom layers will have different responses to external electric fields, even though the two layers are in the same polarization state.
Namely, there is a difference in the energy barriers between the top and the bottom layers with the same type of polarizations (pointing the same direction) to be flipped.
This difference is due to that the $A$ atoms close to and away from the interface behave differently respect to the fields, which leads to a layer-selective flipping for
the polarizations under electric fields. Taking the state C$_1$ ($\frac{\uparrow \uparrow}{\uparrow \uparrow}$) as an example, under an electric field along $-z$, the
polarizations in the top layer are easier to be flipped than those in the bottom layer. Thus, the polarizations of the top layer will be flipped prior to those in the
bottom layer. Further increasing the electric field may drive the polarizations in the bottom layer to be flipped. Moreover, the structures of bilayers with
neighboring values of $P_{tot}^{2L}$ under the electric fields differ only by a half-layer. Therefore, the unique combination of the polarization and layer
DoF gives rise to a layer-selective half-layer-by-half-layer flipping for the transforming between neighboring polarization states (see Fig.~\ref{fig1}b).

We now come to material realization of the concept of polarization multistates. We perform density-functional theory (DFT) calculations for bilayers and trilayers of
both CuInP$_2$S$_6$ and CuCrP$_2$S$_6$ using the Vienna Ab initio Simulation Package~\cite{kresse1996}. The reason for choosing these two systems is that for
both the AFE ordering has a lower energy than the FE ordering. However, the energy difference between the AFE and FE states for them are significantly
different: $\Delta E$ = 24 meV for the CuInP$_2$S$_6$ monolayer and 68 meV for CuCrP$_2$S$_6$, respectively. The former can be tuned into FE via vdW-type interface
interactions. While the latter remains AFE as interfaced with a transition metal dichalcogenide monolayer~\cite{yu2021interface}. The pseudopotentials were
constructed by the projector augmented wave method~\cite{bloechl1994,kresse1999}. A 9 $\times$ 9 $\Gamma$-centered $k$-mesh was used to sample the 2D Brillouin
zone (BZ) and a plane-wave energy cutoff of 400 eV was used for structural relaxations and electronic structure calculations. vdW dispersion forces between the
adsorbate and the substrate were accounted for through the DFT-D3 method~\cite{grimme2010consistent}.

\begin{figure}[h]
  \includegraphics[width=0.95\linewidth]{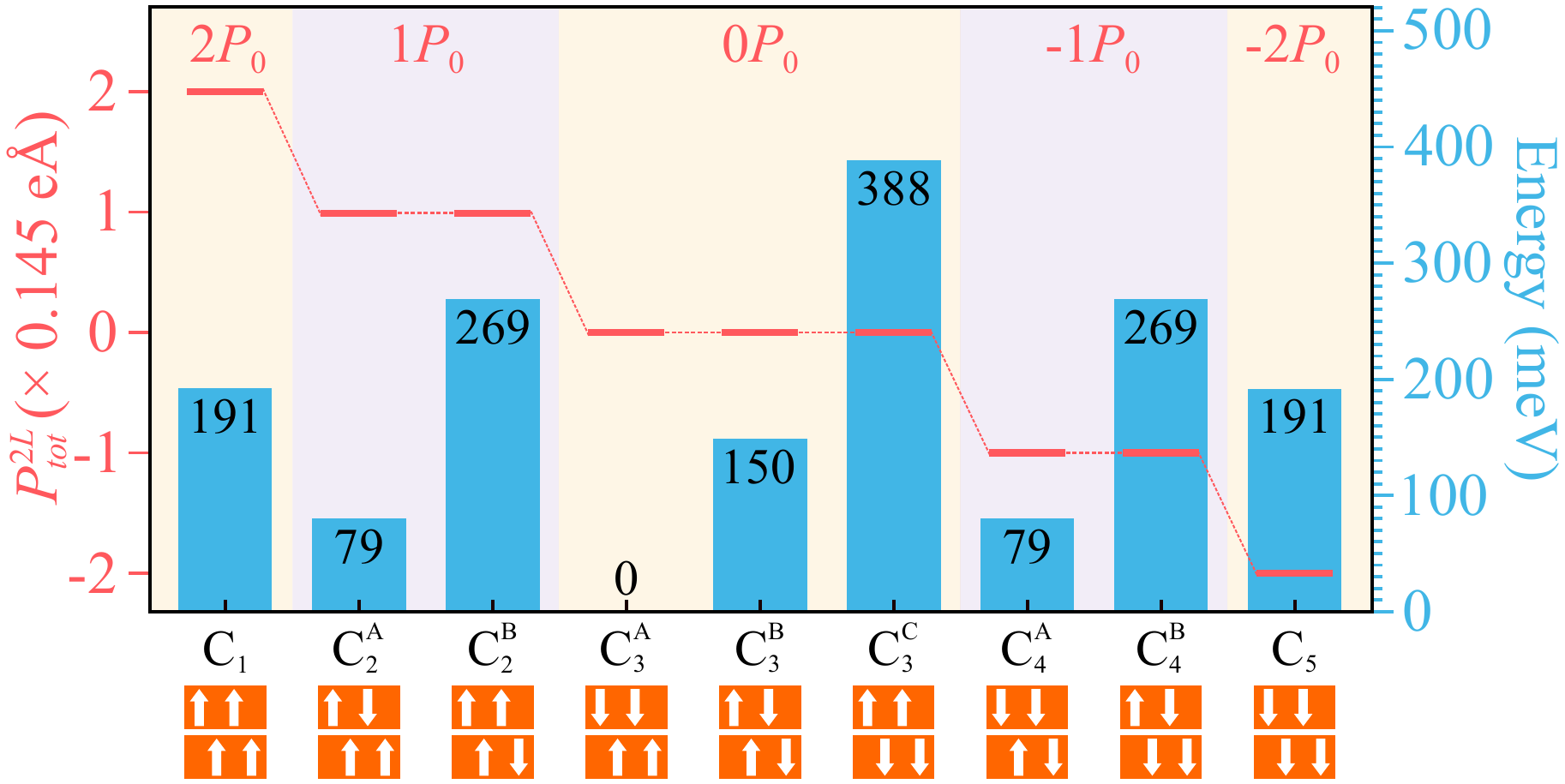}
\caption{DFT-derived polarizations (red line) and energies (blue bars) of a CuInP$_2$S$_6$ bilayer with quintuple polarization states. $P_{tot}^{2L}$ is shown in the unit
of that ($P_0$) for the FE CuInP$_2$S$_6$ monolayer. The energy of the ground state ($C_{3}^A$) is taken as the reference for calculating the energies.}
 \label{fig2}
\end{figure}

Figure~\ref{fig2} shows the DFT results for a CuInP$_2$S$_6$ bilayer. The stacking order is exactly the same as that in the bulk phase of CuInP$_2$S$_6$, in which
neighboring layers are in the H-stacking. We have performed calculations for other stackings, which suggest that they energetically unfavorable. The groud state structure
is the "0" state which was also recently found for the bilayers of CuBiP$_2$Se$_6$ and AgBiP$_2$Se$_6$~\cite{reimers2018van,tawfik2018van,shang2022stacking}. For
CuInP$_2$S$_6$, C$_3^A$ has the lowest energy structure with an intra-layer FE ordering and an inter-layer AFE coupling. Whereas for CuCrP$_2$S$_6$, C$_3^B$ is the
ground state structure, in which the intra-layer AFE ordering of the monolayer is preserved (see Figs.~S7 and S8). The reason for the difference is that the energy
difference between the intra-layer FE and AFE orderings is small for the CuInP$_2$S$_6$ monolayer ($\sim$ 24 meV \cite{yu2021interface}). Therefore, their relative
stability can be modulated by vdW-type layer interactions. In contrast, it is 68 meV for a CuCrP$_2$S$_6$ monolayer, much larger than the energy gain due to the vdW-type
interface interactions.  There are only small changes in the displacements of Cu almost in the bilayers compared to those in freestanding monolayers (less than 8\%).
Therefore, the polarizations derived from our calculations for the quintuple states are roughly quantized in the unit of that for the freestanding systems [see Fig.~\ref{fig2}].

Another trend in the energies of the structures is that the configurations with "head-to-head" polarizations have a lower energy than their cousins with "tail-to-tail"
polarizations. Fig.~\ref{fig2} shows that for the "0" states, C$_3^A$ ($\frac{\downarrow \downarrow}{\uparrow \uparrow}$) is energetically much lower than C$_3^C$
($\frac{\uparrow \uparrow}{\downarrow \downarrow}$). While these two are respectively lower and higher than the "2" state (C$_1$) as well as the "-2" state (C$_5$). This
trend can be understood with interlayer band hybridizations related to the structural differences between them. For C$_3^A$, all the Cu atoms stay close tothe interfaces.
However, for C$_1$ (C$_{5}$), there are only half of them close to the interface. Whereas for C$_3^C$, all the Cu atoms are further away from it (see Figs.~S7 and S8).
Therefore, the interlayer band hybridizations in C$_3^A$ are stronger than those in C$_1$ (C$_5$) and the latter has stronger interlayer band hybridizations than C$_3^C$.
Such a trend in band hybridizations can be seen from the charge density difference, which shows that C$_3^A$ and C$_3^C$ have the largest and smallest charge
redistributions, respectively. For the "1" ("-1") state, C$_2^A$ (C$_4^A$) has a lower energy than C$_2^B$ (C$_4^B$), which can also be understood with the above argument.

\begin{figure}
  \includegraphics[width=0.9\linewidth]{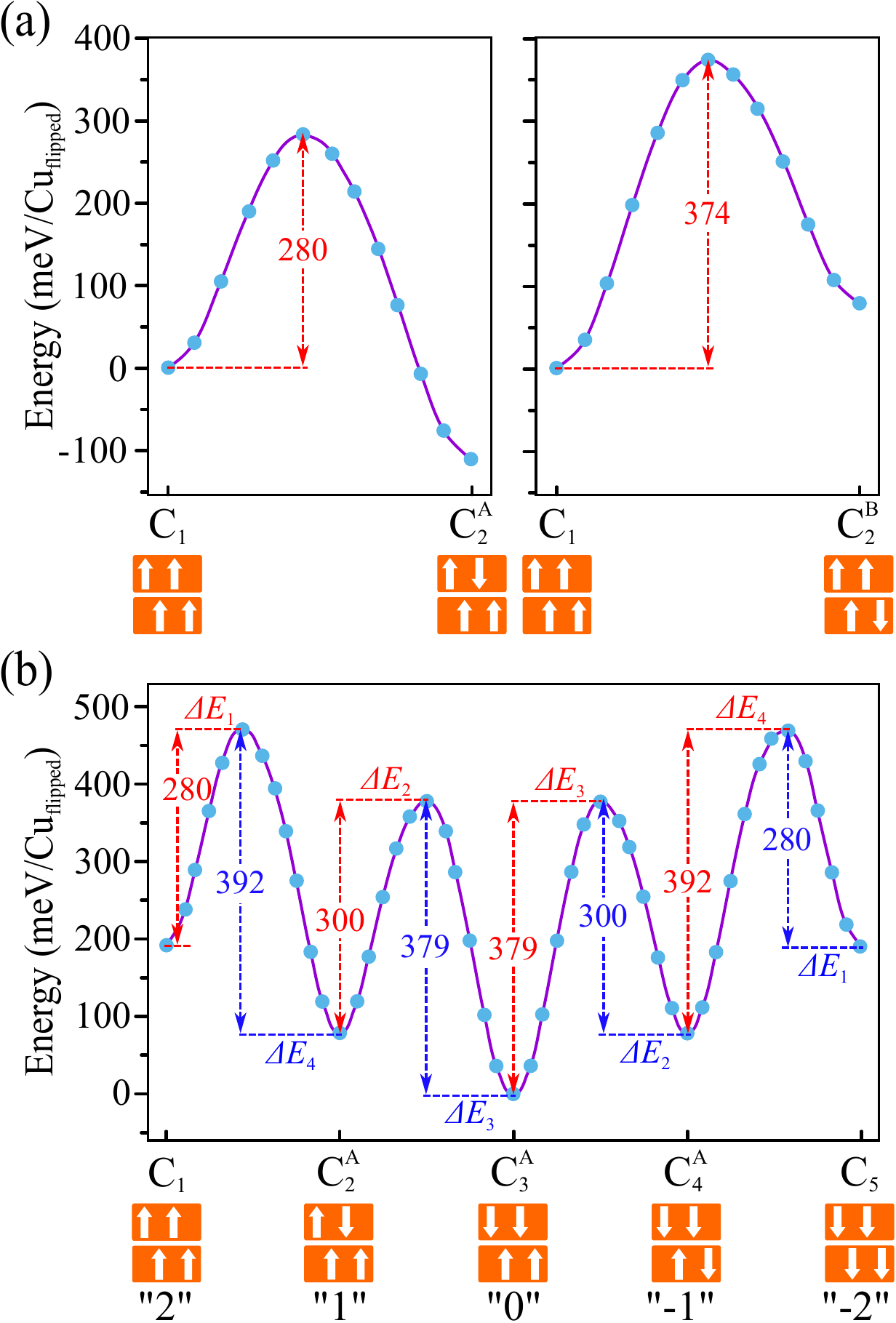}
\caption{Layer-selective half-layer-by-half-layer flipping in a CuInP$_2$S$_6$ bilayer. (a) Comparison of barriers for flipping polarizations in different layers as
transforming from "2" to "1". Here, the total energy of the "2" state ($C_1$) is used as the reference. (b) The kinetic pathways for the transformings between
neighboring polarization states of the bilayer. $\Delta E_i$ label the energy barriers, whose values are shown in red (along the path from "2" to "-2") or blue
(along the path from "-2" to "2").}
 \label{fig3}
\end{figure}

We now discuss transformings of the multiple polarization states in ABP$_2$X$_6$ bilayers.  Figure~\ref{fig3} shows the kinetic pathways of state transformings for a CuInP$_2$S$_6$
bilayer using the climbing image nudged elastic band (CI-NEB) method~\cite{henkelman2000climbing,henkelman2000improved}. One can see that all the polarization states are energy
local minimums, suggesting that they can be stable. The transformings between neighboring states are completed under the mechanism of the layer-selective half-layer-by-half-layer
flipping. For the "2" state, the top layer will be firstly transformed into the AFE state by overcoming an energy barrier of about 280 meV (see Fig.~\ref{fig3}a and $\Delta E_1$ in
Fig.~\ref{fig3}b) such that the bilayer flows into the "1" state. This transforming is confirmed by examing the barriers for all the possible pathways. For instance,
the barrier for the same transforming in the bottom layer is about 94 meV higher than that for this pathway (right panel of Fig.~\ref{fig3}a). Then, an AFE-FE transforming in the
top layer that overcomes an energy barrier of 300 meV ($\Delta E_2$) leads to that the bilayer switches from the "1" state to the "0" state. The transformings from the "0" state
to the "-1" state and later from the "-1" state to the "-2" state are achieved by overcoming energy barriers of about 379 meV ($\Delta E_3$) and 392 meV ($\Delta E_4$), respectively,
by flipping half of the polarizations in the bottom layer. The energy barrier for the transformings from the "-2" state to the "2" state are the same as the above barriers. Such a
symmetry in the barrier is related to the structural symmetry, that is, the structures shown in with Fig.~\ref{fig3}b opposite polarizations can be transformed into each other by an
inversion symmetry. Likewise, the layer-selective half-layer-by-half-layer flippings can be seen during the transformings between the states for other ABP$_2$X$_6$ bilayers, although they may have a different configuration for the "0" state.  The mechanism is also seen in the calculations of domain walls and their motions.

\begin{figure}
  \includegraphics[width=0.9\linewidth]{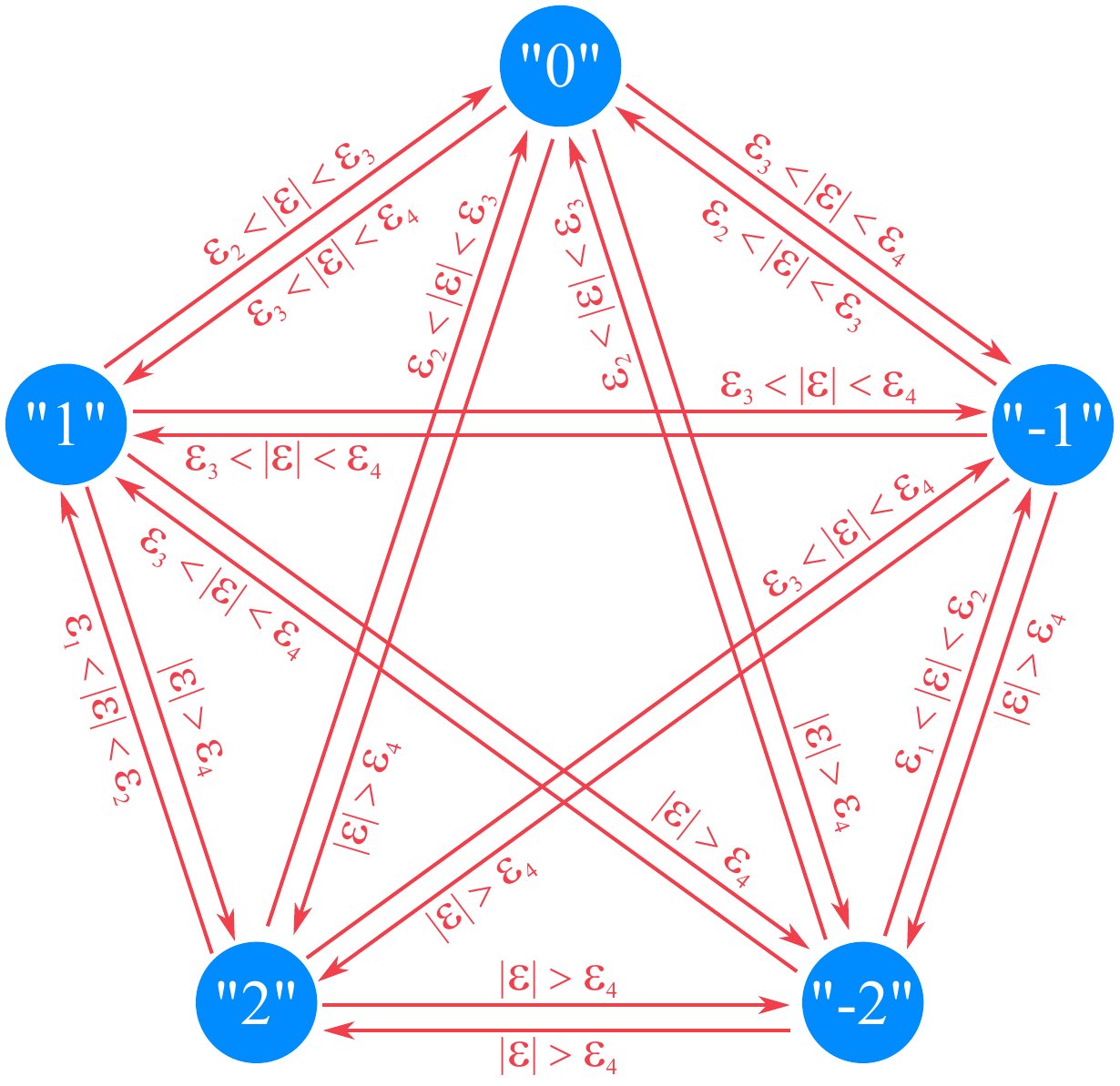}
\caption{Schematical illustration of manipulating the polarization multistates by controlling external electric field. $\varepsilon_i$ denote the magnitude of the electric field
that needed to overcome the energy barrier $\Delta E_i$ shown in Fig.~\ref{fig3}.}
 \label{fig4}
\end{figure}

Figure~\ref{fig3} also shows a feature that the energy barrier between neighboring states increases as the state transforming from the "2" state to the "-2" state. If we label the
critical electric field that needed to overcome the barrier $\Delta E_i$ as $\varepsilon_i$. Then, one can expect that $\varepsilon_1 < \varepsilon_2 < \varepsilon_3 < \varepsilon_4$.
This trend allows switching of any pair of the polarization states by applying an appropriate electric field. For instance, suppose that the bilayer is in the "2" state, an antiparallel
electric field that is larger than $\varepsilon_1$ but smaller than $\varepsilon_2$, i.e., $\varepsilon_1 < \varepsilon < \varepsilon_2$, will transform it into the "1" state. When
$\varepsilon_2 < \varepsilon < \varepsilon_3$, the system will be directly driven into the "0" state without stopping at the "1" state since the electric field will overcome the
barriers $\Delta E_1$ and $\Delta E_2$. Likewise, an antiparallel electric field with $\varepsilon > \varepsilon_4$ will drive the bilayer directly into the "-2" state. Fig.~\ref{fig4}
schematically shows how to control the electric field for the all the possible transformings of the polarization states.  Moreover,  our calculations of ABP$_2$X$_6$ bilayers under  electric fields along the $z$ axis show that the energy barriers can be effectively modified by electric fields. Therefore, the quintuple polarization states in the bilayers
are switchable and controllable. 

We have performed calculations for trilayers of the two systems, for which septuple polarization states are expected (see Figs.~S20 and S21). However, the differences between
neighboring barriers decrease dramatically. For the CuInP$_2$S$_6$ trilayer, there is a difference of 3 meV between $\Delta E_3$ and $\Delta E_4$. For CuCrP$_2$S$_6$, the
minimum in the barrier gradient is about 23 meV, i.e. $\Delta E_3$ - $\Delta E_2$, which is less than half of that for the bilayer. So, the multistates may be indistinguishable
as the thickness of the films grows and the bilayers appear as a better candidate than thicker systems for the multistate operations.

We further investigate the effect of substrates on the polarization multistates in the above bilayers, for which graphene is chosen since it has been widely used in FE devices.
Overall, the graphene substrate tends to reduce the barrier gradient. For CuInP$_2$S$_6$, the difference $\Delta E_3$ and $\Delta E_4$ is reduced from 15 meV to
about 10 meV. This energy window gives roughly 0.01 V/nm for the electric field. For CuCrP$_2$S$_6$, it is about 0.02 V/nm. Previous studies show that the FE switching can be
achieved under a gate voltage of 8 V for thin films of CuInP$_2$S$_6$ with 400 nm, i.e., 0.02 V/nm for the electric field. Therefore, the polarization multistates are expected
to be accessible and distinguishable by current experimental techniques.

In summary, we have revealed switchable polarization multistates in vdW multilayer systems built from stacking AFE ABP$_2$X$_6$ monolayers with out-of-plane polarization.
By making use of the polarization and layer DoF,  one can build quintuple and septuple states for their bilayers and trilayers, respectively. We have demonstrated the concept in thin
films of CuInP$_2$S$_6$ and CuCrP$_2$S$_6$ by performing first-principles and CI-NEB calculations. Our calculations reveal a unique layer-selective half-layer-by-half-layer flipping
of polarization and an increasing barrier height as the polarization difference between states increases, suggesting the multiple states can be transformed into each other by
applying appropriate electric fields. We have further investigated the effects of substrate and thickness.  We anticipate that polarization multistates can also be found in homobilayers and heterostructures of other two-dimensional AFE materials.  Such polarization multistates not only enables significant improvement in the storage density of FE memory devices, but also allows designing new concept
devices that go beyond the bistability paradigm.

\begin{acknowledgments}
This work was supported by the National Natural Science Foundation of China (Grants No. 12174098, No. 11774084, No. U19A2090 and No. 91833302).
\end {acknowledgments}

\bibliography{references}
\bibliographystyle{apsrev4-1}

\end{document}